\documentstyle[12pt]{article}
\advance\textheight by \topskip
\begin{document}
\input epsf

\thispagestyle{empty}
\begin{flushright}
{SU-ITP-95-25} \\ gr-qc/9601005 \\ January 3, 1995 \\
\end{flushright}
\begin{center}
{\Large\bf NONPERTURBATIVE AMPLIFICATION OF \\ \vskip
0.2cm INHOMOGENEITIES IN A SELF-REPRODUCING\\ \vskip
0.5cm UNIVERSE} \vskip 1.1cm {\bf
Andrei Linde} \\ \vskip
0.01cm Department of Physics, Stanford University, Stanford CA
94305-4060, USA \vskip 0.1cm {\bf Dmitri Linde}\\ \vskip 0.01cm California
Institute of
Technology, Pasadena, CA 91125, USA \vskip 0.1cm {\bf Arthur
Mezhlumian}\\ \vskip
0.01cm Department of Physics, Stanford University, Stanford CA
94305-4060, USA\\
\end{center}

\vskip .3 cm {\centerline{\large ABSTRACT}}
\begin{quotation} \vskip -0.3cm
We investigate the distribution of
energy density in a stationary self-reproducing inflationary
universe.  We show that the main fraction of volume of the universe
in a state with a given density $\rho$ at any given moment of proper
time $t$ is concentrated near the centers of deep exponentially wide
spherically symmetric wells in the density distribution. Since this statement
is very surprising and counterintuitive, we perform our investigation  by three
different analytical
methods to verify our conclusions, and then  confirm our analytical results
by computer
simulations.  If one assumes that we are typical observers living in the
universe at a given moment of time, then our results may imply that we
should live near the center of a deep and exponentially large void, which we
will call infloid.
Validity of this particular interpretation of our results is not quite
clear since it depends on the as-yet unsolved problem of measure in
quantum cosmology. Therefore at the moment we would prefer to consider our
results simply as a demonstration of nontrivial
properties of the hypersurface of a given time in the fractal
self-reproducing universe, without making any far-reaching
conclusions concerning the structure of our own part of the universe.   Still
we believe that our results may be of some importance since they demonstrate
that nonperturbative effects in quantum cosmology, at least in principle, may
have significant observational consequences, including an apparent violation of
the
Copernican principle.
\end{quotation}

\newpage

\section{Introduction} \label{Introduction}
According to the Copernican principle, the only special thing about
the Earth is that we are living here. We are not at the center of the
universe, as people thought before. This point of view is reflected
also in the so-called cosmological principle, which asserts that our
place in the universe is by no means special and that the space
around us has to be homogeneous and isotropic after smoothing over
small lumps of matter.  This principle lies in the foundation of
contemporary cosmology \cite{Peebles} since it has not only definite
philosophical appeal but also an apparent observational confirmation
by a host of data on large scale structure of the universe. However,
theoretical interpretation of this principle is usually based on the
big bang picture of the universe and its evolution, inherently related
to simple geometry of Friedmann-Robertson-Walker type. The only
theoretical justification of homogeneity and isotropy of the universe
which is known to us at present is based on inflationary cosmology.
But this theory simultaneously with explaining why our universe
locally looks so homogeneous predicts that on an extremely large scale
the universe must be extremely inhomogeneous \cite{MyBook}. Thus,
after providing certain support to the cosmological principle,
inflationary theory eventually removes it as having only limited
validity.  But until very recently we did not suspect that inflation
may invalidate the Copernican principle as well, since there is
nothing about inflation which would require us to live in the center
of the universe.

The situation became less obvious when we studied the global structure
of inflationary universe in the chaotic inflation scenario, and found
that according to a very wide class of inflationary theories, the main
fraction of volume of the universe in a state with a given density
$\rho$ at any given moment of time $t$ (during or after inflation)
should be concentrated near the centers of deep exponentially wide
spherically symmetric wells in the density distribution
\cite{LLMcenter}.  This result is based on investigation of
nonperturbative effects in the theory of a fractal self-reproducing
universe in the chaotic inflation scenario
\cite{Linde_eternal}.\footnote{Self-reproduction of the universe is
  possible in the new inflationary theory as well
  \cite{Vil_reproduction}, but as we will see, in this theory the
  effect which we are going to discuss is negligibly small.}

Observational implications of this result depend on its
interpretation.  If we assume that we live in a part which is typical,
and by ``typical'' we mean those parts of the universe which have the
greatest volume with other parameters (time and density) being equal,
then our result implies that we should live near the center of one of
the wells in the density distribution. There should be many such wells
in the universe, but each of them should be exponentially wide.  In
what follows we will call these wells ``infloids.''  An observer
living near the center of an infloid will see himself ``in the center
of the world,'' which would obviously contradict the Copernican
principle.

One should clearly distinguish between the validity of our result and
the validity of its interpretation suggested above. Even though the
effect by itself is rather surprising we think that it is correct. We
verified its validity by three independent analytical methods, as well
as by computer simulations.  Meanwhile the validity of its
interpretation is much less clear.  The main problem is related to the
ambiguity in the choice of measure in quantum cosmology \cite{LLM}.
There are infinitely many domains with similar properties in a
self-reproducing inflationary universe.  When we are trying to compare
their volumes, we are comparing infinities.  The results of this
comparison depend on the choice of the regularization procedure. The
prescription that we should compare volumes at a given time $t$ in
synchronous coordinates is intuitively appealing, but there exist
other prescriptions which lead to different conclusions
\cite{LLMcenter,LLM,Vil_predict_1,LMregul}.  Until the interpretation
problem is resolved, we will be unable to say for sure whether
inflationary cosmology actually predicts that we should live in a
center of a spherically symmetric well.  Still this possibility is so
interesting that it deserves a detailed investigation even at our
present, admittedly rather incomplete level of understanding of
quantum cosmology. This is the main purpose of our paper.

In Section \ref{Typical} we will give a short review of the theory of
self-reproducing universe in the chaotic inflation scenario and
discuss which type of phenomena should be called typical in such
universe. Then we will describe two approaches to the problem of
estimating the typical magnitude of the quantum fluctuations under the
volume weighted measure. The first is based on counting the balance of
probability factors. The second is based on the investigation of the
probability distribution $P_p(\phi,t)$.  This distribution describes
the portion of the physical volume of the universe which contains the
field $\phi$ at the time $t$. According to \cite{LLM}, this
distribution rapidly approaches a stationary regime, where the portion
of the physical volume of the universe containing the field $\phi$
becomes independent on time.  Investigation of this distribution in
Section \ref{Nonpert_Effects} will allow us to derive our result in a
different way.  In Section \ref{VolumeRoll} we will develop a path
integral approach to the investigation of $P_p(\phi,t)$. The new
method provides another way to confirm our results.  However, this
method is interesting by itself. It gives us a new powerful tool for
investigation of the global structure of the self-reproducing
universe, which may be useful independently of existence of the effect
discussed in this paper. In Section \ref{Simul} we will describe
computer simulations which we used to verify our analytical results.
Only then, after we make sure that our rather counterintuitive results
are actually correct, we will describe their possible interpretation
and their observational consequences.  In Section \ref{Infloids} we
will describe the structure of infloids, their evolution after the end
of inflation and their observational manifestations. In Section
\ref{Interpretation} we will discuss our results, ambiguities of their
interpretation, and formulate our conclusions. In Appendix we present generalization of our results for different time parametrizations.


\section{The Self-Reproducing Universe}
\label{Typical}

Let us consider the simplest model of chaotic inflation based on the
theory of a scalar field $\phi$ minimally coupled to gravity, with the
effective potential $V(\phi)$. If the classical field $\phi$ (the
inflaton field) is sufficiently homogeneous in some domain of the
Universe, then its behavior inside this domain is governed by the
equations

\begin{equation}\label{phi_motion}
\ddot\phi + 3H\dot\phi = -V'(\phi)\ ,
\end{equation}

\begin{equation}\label{Hubble_complete}
H^2 + \frac{k}{a^2} =
\frac{8\pi}{3M^2_p}\,
\left(\frac{1}{2} \dot\phi^2 + V(\phi)\right) \ .
\end{equation}
Here $H={\dot a}/a,\, a(t)$ is the scale factor of the Universe,
$k=+1, -1,$ or $0$ for a closed, open or flat Universe, respectively.
$M_p$ is the Planck mass, which we will put equal to one in the rest
of the paper.

Investigation of these equations has shown that for many potentials
$V(\phi)$ (e.g., in all power-law $V(\phi)\sim \phi^n$ and exponential
$V(\phi)\sim e^{\alpha \phi}$ potentials) there exists an intermediate
asymptotic regime of slow rolling of the field $\phi$ and
quasi-exponential expansion (inflation) of the Universe \cite{MyBook}.
At this stage, which is called inflation, one can neglect the term
$\ddot\phi$ in (\ref{Hubble_complete}), as well as the terms
$\frac{k}{a^2}$ and $\frac{4\pi}{3}\dot\phi^2$ in
(\ref{Hubble_complete}).  Therefore during inflation

\begin{equation}\label{slow_roll}
 H = \sqrt{8\pi V\over 3}\ ,
{}~~~~~~\dot\phi = -{V'(\phi)\over 3 H}\ .
\end{equation}
 In the theories $V(\phi)\sim \phi^n$ inflation ends at
$\phi = \phi_e$, where $\phi_e \sim 10^{-1}\, n$.

Inflation stretches all initial inhomogeneities. Therefore, if the
evolution of the Universe were governed solely by classical equations
of motion, we would end up with an extremely smooth Universe with no
primordial fluctuations to initiate the growth of galaxies.
Fortunately, new density perturbations are generated during inflation
due to quantum effects. The wavelengths of all vacuum fluctuations of
the scalar field $\phi$ grow exponentially in the expanding Universe.
When the wavelength of any particular fluctuation becomes greater than
$H^{-1}$, this fluctuation stops oscillating, and its amplitude
freezes at some nonzero value $\delta\phi (x)$ because of the large
friction term $3H\dot{\phi}$ in the equation of motion of the field
$\phi$\@. The amplitude of this fluctuation then remains almost
unchanged for a very long time, whereas its wavelength grows
exponentially. Therefore, the appearance of such a frozen fluctuation
is equivalent to the appearance of a classical field $\delta\phi (x)$
that does not vanish after averaging over macroscopic intervals of
space and time.

Because the vacuum contains fluctuations of all wavelengths, inflation
leads to the creation of more and more perturbations of the classical
field with wavelengths greater than $H^{-1}$\@. The average amplitude
of such perturbations generated during a time interval $H^{-1}$ (in
which the Universe expands by a factor of e) is given by

\begin{equation}\label{phi_fluctuation}
\left| \delta\phi(x)\right| = \frac{H}{2\pi}\ .
\end{equation}
The phases of each wave are random. It is important also that quantum
fluctuations occur independently in all domains of inflationary
universe of a size greater than the radius of the event horizon
$H^{-1}$.  Therefore, the sum of all waves at any given region of a
size $O(H^{-1})$ fluctuates and experiences Brownian jumps in all
directions in the space of fields. The standard way of description of
the stochastic behavior of the inflaton field during the slow-rolling
stage is to coarse-grain it over separate domains of radius $H^{-1}$
(we will call these domains ``$h$-regions'' \cite{ArVil,Nambu}, to
indicate that each of them has the radius coinciding with the radius
of the event horizon $H^{-1}$\@) and consider the effective equation
of motion of the long-wavelength field \cite{Star,Linde_eternal}:

\begin{equation} \label{SDE}
\frac{d}{dt} \, \phi = - \frac{V'(\phi)}{3H(\phi)} +
\frac{H^{3/2}(\phi)}{2\pi} \, \xi(t) \ ,
\end{equation}
Here $\xi(t)$ is the effective white noise generated by quantum
fluctuations.

Let us find the critical value $\phi_*$ such that for $V(\phi) <
V(\phi_*)$ the classical slow roll dominates the evolution of the
inflaton, while for $V(\phi) > V(\phi_*)$ the quantum fluctuations are
more important. Within the characteristic time interval $\Delta t =
H^{-1}$ for values of inflaton near the critical value $\phi_*$ the
classical decrease $\Delta \phi = \dot \phi \, \Delta t$ of the
inflaton, defined through (\ref{slow_roll}), is of the same magnitude
as the typical quantum fluctuation generated during the same period,
given by (\ref{phi_fluctuation}).  After some algebra we get from
equations (\ref{slow_roll}) and (\ref{phi_fluctuation}) the relation
defining $\phi_*$ implicitly:
\begin{equation}
  \label{phi_*_define}
  \frac{3 H^3(\phi_*)}{2 \pi \, V'(\phi_*)}
= H(\phi_*) \, \frac{4\, V(\phi_*)}{V'(\phi_*)} \sim 1 .
\end{equation}

Let us consider for definiteness the theory $V(\phi) = \lambda \phi^4
/4$. In this case equation (\ref{phi_*_define}) yields $\phi_* \sim
{\lambda}^{-1/6} $. One can easily see that if $\phi < \phi_*$, then
the decrease of the field $\phi$ due to its classical motion
$\Delta\phi = 1/2\pi\phi$ is much greater than the average amplitude
of the quantum fluctuations $\delta\phi = \sqrt{\lambda/ 6\pi}\phi^2$
generated during the same characteristic time interval $H^{-1}$. But
for $\phi > \phi_*$, $\delta\phi (x)$ will exceed $\Delta\phi$, i.e.
the Brownian motion of the field $\phi$ will become more rapid than
its classical motion.  Because the typical wavelength of the
fluctuations $\delta\phi (x)$ generated during this time is $H^{-1}$,
the whole Hubble domain after the time $H^{-1}$ becomes effectively
divided into $e^3$ $h$-regions, each containing almost homogeneous
(but different from each other) field $\phi -
\Delta\phi+\delta\phi$\@.

In almost half of these domains (i.e. in $e^3/2 \sim 10$ $h$-regions)
the field $\phi$ grows by $|\delta\phi(x)|-\Delta\phi \approx
|\delta\phi (x)| = H/2\pi$, rather than decreases. During the next
time interval $\Delta t = H^{-1}$ the field grows again in the half of
the new $h$-regions. Thus, the total number of $h$-regions containing
growing field $\phi$ becomes equal to $(e^3/2)^2 = e^{2\,(3 - \ln
  2)}$\@.  This means that until the fluctuations of field $\phi$ grow
sufficiently large, the total physical volume occupied by permanently
growing field $\phi$ (i.e. the total number of $h$-regions containing
the growing field $\phi$) increases with time like $\exp[(3 - \ln
2)\,Ht]$\@. This leads to the self-reproduction of inflationary
domains with $\phi > \phi_*$ in the chaotic inflation scenario
\cite{Linde_eternal}.

Note, that the greater is the value of the effective potential, the
greater is the rate of exponential expansion of the universe. As a
result, the main growth of the total volume of the universe occurs due
to exponential expansion of the domains with the greatest possible
values of the Hubble constant $H = H_{\rm max}$
\cite{Linde_eternal,LLM}. In some models there is no upper bound to
the value of $H$ \cite{BD,BL}.  However, in the simplest versions of
chaotic inflation based on the Einstein theory of gravity there are
several reasons
to expect that there exists an upper bound for the rate of inflation \cite{LLM,VilMediocre,GBLinde}.

In what follows we will assume that there is an upper bound $H_{\rm
  max}$ on the value of the Hubble constant during inflation. For
definiteness we will assume that $H_{\rm max} = \sqrt{8\pi/ 3}$, which
corresponds to the Planck boundary $V(\phi_p) = 1$. This is a rather
natural assumption for chaotic inflation. However, one should note
that in some models $H_{\rm max}$ may be much smaller. In particular,
in the new inflation scenario $H_{\rm max} = \sqrt{8\pi V(0)/ 3}$ \,
is many orders of magnitude smaller than 1.

The independence of the subsequent evolution of the $h$-region on its
previous history, the dominance of the domains where the inflaton
field energy grows rather than decreases in the volume weighted measure
and the upper bound for the energies at which the inflation can
proceed are the three main features inherent to many models of
inflation. When all these features are present the evolution of the
inflationary universe as a whole approaches regime which we called {\it
  global stationarity} in \cite{LLM}. This stage is characterized by the
stability of the distribution of regions with various local values of
energy density and other parameters, while the number of such regions
grows exponentially with a constant coefficient, proportional to the
maximal possible rate of inflation $\lambda_1 = d_{\rm fr}\, H_{\rm
  max}$. Here $d_{\rm fr}$ is a model dependent fractal dimension of
the classical space \cite{ArVil,LLM}, which is very close to $3$ for small
coupling constants of the inflaton field.

The new picture of the universe is extremely unusual, and it may force us to reconsider our definition of what is typical and what is not. In particular, the standard theory of the large scale structure of the universe is based on the assumption that a typical behavior of the scalar field at the last stages of inflation is described by equations (\ref{slow_roll}), (\ref{phi_fluctuation}). 
This is indeed the case if one studies a single branch of inflationary universe beginning at $\phi \ll \phi_*$. However, if one investigates the global structure of the universes at all $\phi$ and tries to find the   typical behavior of {\it all} inflationary domains {\it with a volume weighted measure}, the result may appear to be somewhat different.

\section{Stationary Inflation and Nonperturbative Effects}
\label{Nonpert_Effects}

 Suppose that we have one
inflationary domain of initial size $H^{-1}$, containing scalar field
$\phi > \phi_*$. Let us wait 15 billion years (in synchronous time $t$
in each part of this domain) and see what are the typical properties
of those parts of our original domain which at the present moment have
some particular value of density, e.g. $\rho =10^{-29} \, {\rm g}
\cdot {\rm cm}^{-3}$. The answer to this question
proves to be rather unexpected.

This domain exponentially expands, and becomes divided into many new
domains of size $H^{-1}$, which evolve independently of each other. In
many new domains the scalar field decreases because of classical
rolling and quantum fluctuations. The rate of expansion of these
domains rapidly decreases, and they give a relatively small
contribution to the total volume of those parts of the universe which
will have density $10^{-29} \, {\rm g} \cdot {\rm cm}^{-3}$ 15 billion
years later. Meanwhile those domains where quantum jumps occur in the
direction of growth of the field $\phi$ gradually push this field
towards the upper bound where inflation can possibly exist, which is
presumably close to the Planck boundary $V(\phi_p) \sim 1$. Such
domains for a long time stay near the Planck boundary, and
exponentially grow with the Planckian speed. Thus, the longer they
stay near the Planck boundary, the greater contribution to the volume
of the universe they give.

However, the domains of interest for us eventually should roll down
and evolve into the regions with density $10^{-29} \, {\rm g} \cdot
{\rm cm}^{-3}$. Thus, these domains cannot stay near the Planck
boundary for indefinitely long time, producing new volume with the
Planckian speed. However, they will do their best if they stay there
as long as it is possible, so that to roll down at the latest possible
moment. In fact they will do even better if they stay near the Planck
boundary even longer, to save time for additional rapid inflation, and
then rush down with the speed exceeding the speed of classical
rolling.  This may happen if quantum fluctuations coherently add up to
large quantum jumps towards small $\phi$. This process is dual to the
process of perpetual climbing up, which leads to the self-reproduction
of inflationary universe.

Of course, the probability of large quantum jumps down is
exponentially suppressed. However, by staying longer near the Planck
boundary inflationary domains get an additional exponentially large
contribution to their volume.  These two exponential factors compete
with each other to give us an optimal trajectory by which the scalar
field rushes down in those domains which eventually give the leading
contribution to the volume of the universe. From what we are saying it
should be clear that the quantum jumps of the scalar field along such
optimal trajectories should have a greater amplitude than their
regular value ${H\over 2\pi}$, and they should preferably occur
downwards. As a result, the energy density along these optimal
trajectories will be smaller than the energy density of their lazy
neighbors which prefer to slide down without too much of jumping.
This creates wells in the distribution of energy density, which we
called infloids \cite{LLMcenter}.

Suppose that the extra time interval spent at highest energies is
$\tilde{\Delta t}$\@. Then we win the volume by factor of
$\exp\left(d_{\rm fr} H_{\rm max} \tilde{\Delta t} \right)$\@.
However, to compensate for the lost time the inflaton field $\phi$ has
to jump at least once (let us say, when it reaches the value $\phi$)
with the amplitude $ \tilde{\delta \phi} = n(\phi) \, H(\phi)/2 \pi$
such that it covers in one jump the distance which would otherwise
require time $\tilde{\Delta t}$ to slowly roll down:

\begin{equation}\label{deltat}
 \tilde{\Delta t}(\phi) =
\frac{\tilde{\delta \phi}}{\dot{\phi}} = \frac{ n(\phi)
\frac{H(\phi)}{2 \pi}}{\dot{\phi}} = n(\phi) \,
\frac{4 V(\phi)}{V'(\phi)} \ ,
\end{equation}
where we introduced the factor $n(\phi)$ by which the jump is
amplified, i.e. by which it is greater than the standard jump
$H(\phi)/2 \pi$. The probability of such jump is suppressed by the
factor $\exp\left(-\frac{1}{2} \, n^2(\phi)\right)$\@.  The leading
contribution to the volume of the universe occurs due to the jumps
which maximize the volume weighted probability:

\begin{equation}\label{maximize}
 P \sim \exp{\left(d_{\rm fr} H_{\rm max} \,
\tilde{\Delta t}(\phi) - \frac{1}{2} n^2(\phi)\right)}
= \exp{\left(d_{\rm fr} H_{\rm max} \,
n(\phi)\, \frac{4 V(\phi)}{V'(\phi)} - \frac{1}{2} n^2(\phi)\right)}  \  .
\end{equation}
Maximizing with respect to $n(\phi)$ gives the amplification factor as
a function of the location of the jump on the inflaton trajectory:

\begin{equation}\label{amplification}
 n(\phi) = 4 \, d_{\rm fr} \, H_{\rm max} \, \frac{V(\phi)}{V'(\phi)} \ .
\end{equation}
In fact, we have found \cite{LLMcenter} that the typical trajectories
which give the leading contribution to the volume of the universe
consist entirely of such subsequent jumps. In what follows we will
give an alternative derivation of this result. Meanwhile, comparing
with (\ref{phi_*_define}) one immediately sees that $n(\phi) \gg 1$
for $\phi < \phi_*$, since $d_{\rm fr} \sim 3$ and $H_{\rm max} \gg
H(\phi)$ for such values of inflaton field in chaotic inflation.
Therefore, our treatment of these quantum fluctuations as large and
rear quantum jumps is self-consistent.

To avoid misunderstandings one should note that a more accurate definition of
amplification coefficient would be $n(\phi) +1$. Indeed,  in the absence of
nonperturbative effects we would have  $n(\phi) = 0$ since perturbative jumps
occur in both directions with equal probability. The  coefficient $n(\phi)$
relates an additional amplitude of jumps {\it down} to the regular perturbative
amplitude of the jumps in both directions. This subtlety will not be important
for us here since we are interested in the case $n \gg 1$.

 It is interesting that the coefficient of amplification $n(\phi)$ can be
directly related to the ratio of   amplitudes of conventional scalar and tensor
perturbations generated at the same scale at which the jump occurs. The
amplitudes of these perturbations can be written as follows:
\begin{equation}\label{perturbscalar}
A^{pert}_S(\phi) = \left(
\frac{\delta \rho}{\rho} \right)_S =   c_S \, \frac{H^2(\phi)}{2\pi\dot{\phi}}
\  ,~~~~~~~  A^{pert}_T(\phi) = \left(
\frac{\delta \rho}{\rho} \right)_T =   c_T \, \frac{H(\phi)
}{M_p}  \ .
\end{equation}
Here $c_S$ and $c_T$ are some coefficients of the order of unity.
Using these expressions   we can rewrite (\ref{amplification}) for $d_{\rm fr}
\sim 3$ in the form:

\begin{equation}\label{maximal2}
 n(\phi) =  \frac{3c_T}{c_S} \,  \frac{H_{\rm max}
}{M_p} \, \frac{A^{pert}_S(\phi)}{A^{pert}_T(\phi)} \ .
\end{equation}

In the same way as the conventional amplitude of jumps $H/2 \pi$ is
related with the well known perturbations of the background energy
density, the ``nonperturbatively amplified'' jumps which we have just
described are related to the ``nonperturbative'' contribution to
deviations of the background energy density from its average value.
A possible interpretation of this result is that at the length scale
associated with the value of the field $\phi$ there is an additional
nonperturbative contribution to the {\it monopole} amplitude:

\begin{equation}\label{nonperturb}
 A^{nonpert}_S(\phi) = \left(
\frac{3 c_T}{c_S} \, \frac{H_{\rm max} }{M_p} \,
\frac{A^{pert}_S(\phi)}{A^{pert}_T(\phi)} \right) \, A^{pert}_S \ .
\end{equation}

We will discuss the structure of infloids and their possible observational
consequences   in Section \ref{Infloids}. Here we only note that eq.
(\ref{maximal2}) gives a simple tool for understanding of the possible
significance of the effect under consideration. Indeed, in the simplest chaotic
inflation models, such as  the theory ${\lambda\over n}\phi^n$, one has $H_{\rm
max} \sim {M_p}$ and $A^{pert}_S(\phi)\gg  A^{pert}_T(\phi) $; thus one has
$n(\phi)  \gg 1$.
On the other hand, in the versions of chaotic inflation scenario where
inflation occurs near a local maximum of the effective potential (as in the new
inflation models) $H_{\rm max}$ is   many orders of magnitude smaller than
${M_p}$, and therefore the non-perturbative effects discussed above are
negligibly small. Thus, investigation of nonperturbative effects can give us a
rather unexpected possibility to distinguish between various classes of
inflationary models. We will return to this issue in the end of the paper.

\section{Nonperturbative effects and branching diffusion}

  One of the best ways to examine nonperturbative effects is
to investigate the probability distribution $P_p(\phi,t)$ to find a
domain of a given physical volume in a state with a given field $\phi$
at some moment of time $t$.  The distribution $P_p(\phi,t)$ obeys the
following  branching  diffusion equation \cite{Nambu,Mijic,LLM}:

\begin{equation}\label{BranchFP}
 \frac{\partial P_p}{\partial t} =
\frac{\partial }{\partial\phi} \left( \frac{H^{3/2}(\phi)}{2\sqrt{2}\pi}\,
\frac{\partial }{\partial\phi}
\left( \frac{H^{3/2}(\phi)}{2\sqrt{2}\pi} P_p \right) +
\frac{V'(\phi)}{3H(\phi)} \, P_p\right) + 3H(\phi) P_p\ .
\end{equation}

This equation is valid only during inflation, which typically occurs
within some limited interval of values of the field $\phi$: \
$\phi_{min} < \phi < \phi_{max}$. In the simplest versions of chaotic
inflation model $\phi_{min}\equiv \phi_e \sim 1$, where $\phi_e$ is
the boundary at which inflation ends. Meanwhile, as we argued in the
previous section, $\phi_{max}$ is close to the Planck boundary
$\phi_p$, where $V(\phi_p)=1$. To find solutions of this equation one
must specify boundary conditions. Behavior of solutions typically is
not very sensitive to the boundary conditions at $\phi_{e}$; it is
sufficient to assume that the diffusion coefficient (and,
correspondingly, the double derivative term in the r.h.s.\ of equation
(\ref{BranchFP})) vanishes for $\phi < \phi_{e}$ \cite{LLM}. The
conditions near the Planck boundary play a more important role.  In
this paper we will assume, that there can be no inflation at $V(\phi)
> 1$, which corresponds to the boundary condition $
P_p(\phi,t)|_{\phi>\phi_p} = 0$. In the end of the paper we will
discuss possible modifications of our results if $\phi_{max}$ differs
from $\phi_p$.

One may try to obtain solutions of equation (\ref{BranchFP}) in the
form of the eigenfunction series:
\begin{equation}
  \label{Eigenseries}
P_p(\phi,t) = \sum\limits_{s=1}^{\infty} { e^{\lambda_s
t}\, \pi_s(\phi) }
\; \stackrel{t \rightarrow \infty}{\longrightarrow} \;
e^{\lambda_1 t}\, \pi_1(\phi) ,
\end{equation}
where, in the limit of large time $t$, only the term with the largest
eigenvalue $\lambda_1$ survives. The function $\pi_1(\phi)$ in the
limit $t \to \infty$ has a meaning of a normalized {\it
  time-independent} probability distribution (so called invariant
probability density of the branching diffusion) to find a given field
$\phi$ in a unit physical volume, whereas the factor $e^{\lambda_1 t}$
shows the overall growth of the volume of all parts of the universe,
which does not depend on $\phi$ in the limit $t\to \infty$.  This
``ground state'' eigenfunction satisfies the following equation:

\begin{equation} \label{StationaryFP}
\frac{\partial }{\partial\phi} \left( \frac{H^{3/2}(\phi)}{2\sqrt{2}\pi}\,
\frac{\partial }{\partial\phi}
\left( \frac{H^{3/2}(\phi)}{2\sqrt{2}\pi} \pi_1(\phi)  \right) +
\frac{V'(\phi)}{3H(\phi)} \, \pi_1(\phi) \right) + 3H(\phi) \pi_1(\phi)
 = \lambda_1 \,  \pi_1(\phi) \ .
\end{equation}

In the limit when we can neglect the diffusion (second derivative)
term it is easy to solve this equation:

\begin{equation}
  \label{no_diffusion_eigenfunction}
  \pi_1(\phi) = C(\phi_0)\, \frac{3H(\phi)}{V'(\phi)}
\exp\left( - \int\limits_{\phi}^{\phi_0}{
\left[ \lambda_1 \, \frac{3 H(\zeta)}{V'(\zeta)}
- \frac{9 H^2(\zeta)}{V'(\zeta)} \right] d \zeta} \right) ,
\end{equation}
where we chose some starting point $\phi_0$ and the corresponding
normalization constant $C(\phi_0)$ which should match this approximate
solution to the exact one at this point. As before, let us introduce
the fractal dimension of classical space-time through $\lambda_1 =
d_{\rm fr} H_{\rm max}$ (see \cite{ArVil, LLM} for detailed discussion
of the fractal structure of self-reproducing universe). Let us also
introduce the critical value of inflaton $\phi_{\rm fr}$ at which the
no-diffusion approximation for (\ref{StationaryFP}) breaks. Then,
since $H_{\rm max} \gg H(\phi)$ for chaotic inflation, we can rewrite
(\ref{no_diffusion_eigenfunction}) as:

\begin{equation}
  \label{no_diffusion_solution}
  \pi_1(\phi) = C(\phi_{\rm fr})\, \frac{3H(\phi)}{V'(\phi)}
\exp\left( - \int\limits_{\phi}^{\phi_{\rm fr}}{
 d_{\rm fr} \, H_{\rm max} \frac{3 H(\zeta)}{V'(\zeta)}
 d \zeta} \right) .
\end{equation}

Substituting (\ref{no_diffusion_solution}) into (\ref{StationaryFP})
we get the defining relation for the value of inflaton field
$\phi_{\rm fr}$ at which the no-diffusion approximation breaks:

\begin{equation}
  \label{phi_fr_define}
  \lambda_1 \, \frac{9 \, H^5(\phi_{\rm fr})}{
    4 \pi^2 \, \left( V'(\phi_{\rm fr}) \right)^2} \sim 1 \ .
\end{equation}

We can rewrite (the square root of) this relation in a form which
makes the comparison with the definition of the other critical value
$\phi_*$ more apparent:

\begin{equation}
  \label{phi_fr_compare}
  \sqrt{  \frac{d_{\rm fr}H_{\rm max}}{H(\phi_{\rm fr})}} \
  \frac{3 \, H^3(\phi_{\rm fr})}{2 \pi \, V'(\phi_{\rm fr})} \sim 1 \ .
\end{equation}
Comparing (\ref{phi_fr_compare}) with (\ref{phi_*_define}), one finds
 that for all chaotic inflation models $\phi_{\rm
  fr} < \phi_*$ (one can assume self-consistently that $H_{\rm max}
\gg H(\phi_{\rm fr})$ in such models). The value of $\phi_*$ in
(\ref{phi_*_define}) comes from comparing the slow roll rate in a
given $h$-region with the typical amplitude of quantum fluctuations
while considering only the $h$-regions generated locally from the
region which we picked. On the other hand, the value $\phi_{\rm fr}$
comes from comparing the slow roll rate to the typical amplitude of
fluctuations considering all $h$-regions in the whole universe which
happen to have the same value of inflaton field inside. The fact that
the second constraint is more stringent is yet another indication of
the considerably larger magnitude of the quantum fluctuations when we
take into account the whole stationary universe.

In the particular case of the simplest theory with $V(\phi) = \lambda
\phi^4/4$, we have $H = \sqrt{2\pi\lambda /3}\phi^2$,   $\phi_{\rm fr} \sim
\lambda^{-1/8} \ll \phi_*$, and the dependence
of the solution (\ref{no_diffusion_solution}) on $\phi$ is \cite{LLM}:

\begin{equation}\label{SMALLPHI}
 \pi_1(\phi) \sim \phi^{\sqrt{6\pi\over
\lambda} \, d_{\rm fr} \, H_{\rm max} }   \ .
\end{equation}
This is an extremely strong dependence. For example,  for the realistic value
of the coupling constant
$\lambda \sim 10^{-13}$ chosen to fit the observable large scale
structure of the universe one has $d_{\rm fr} \approx 3$. One may assume for
definiteness that $H_{\rm max} = \sqrt{8 \pi / 3}$, corresponding to
inflation with $V(\phi) = 1$ (Planck density). Then one has  an extremely sharp
dependence $\pi_1 \sim
\phi^{1.2\cdot 10^8}$. All surprising results we are going to
obtain are rooted in this effect.  One of the consequences is the
distribution of energy density $\rho$. For example, during inflation
$\rho \approx \lambda \phi^4/4$. Equation (\ref{SMALLPHI}) implies
that the distribution of density $\rho$ is
\begin{equation}\label{SMALLPHI_1}
P_p(\rho) \sim
\rho^{3\cdot10^7} \ .
\end{equation}
Thus at each moment of time $t$ the universe
consists of indefinitely large number of domains containing matter
with all possible values of density, the total volume of all domains
with density $2\rho$ being approximately $10^{10^7}$ times greater
than the total volume of all domains with density $\rho$!

Let us consider now all inflationary domains which contain a given
field $\phi$ at a given moment of time $t$. One may ask the question,
what was the value of this field in those domains at the moment $t -
H^{-1}$ ? In order to answer this question one should add to $\phi$
the value of its classical drift $\dot\phi H^{-1}$ and the amplitude
of quantum jumps $\Delta \phi$. The typical jump is given by $\delta
\phi = \pm H/ 2\pi$. At the last stages of inflation this quantity is
by many orders of magnitude smaller than $\dot\phi H^{-1}$. But in
which sense jumps $\pm H/ 2\pi$ are typical? If we consider any
particular initial value of the field $\phi$, then the typical jump
from this point is indeed given by $\pm H/2\pi$ under the conventional
comoving measure. However, if we are
considering all domains with a given $\phi$ and trying to find all
those domains from which the field $\phi$ could originate back in
time, the answer may be quite different. Indeed, the total volume of
all domains with a given field $\phi$ at any moment of time $t$
depends on $\phi$ extremely strongly: the dependence is exponential in
general case (\ref{no_diffusion_solution}), or a power law with a huge
power, like in the case of $\lambda \phi^4/4$ theory (\ref{SMALLPHI}).
This means that the total volume of all domains which could jump
towards the given field $\phi$ from the value $\phi +\Delta \phi$ will
be enhanced by a large additional factor $ P_p(\phi +\Delta \phi) /
P_p(\phi)$. On the other hand, the probability of large jumps
$\Delta\phi$ is suppressed by the Gaussian factor $\exp\left(- 2 \pi^2
  \, (\Delta\phi)^2/ H^2 \right)$. Thus, under the established
stationary probability distribution the probability of the inflaton
field in a given domain to have experienced a quantum jump
$\Delta\phi$ is given by:

\begin{equation}
  \label{jump_probability}
  P(\Delta \phi) \sim \exp\left(
 d_{\rm fr} \, H_{\rm max} \frac{3 H(\phi)}{V'(\phi)}
 \Delta \phi -  \frac{2 \pi^2 \, (\Delta\phi)^2}{H^2(\phi)}  \right) .
\end{equation}
One can easily verify that this distribution has a sharp maximum at:

\begin{equation}
  \label{jump_size}
  \Delta \phi_{\rm np} = d_{\rm fr} \, H_{\rm max}
  \frac{3 H^3(\phi)}{4 \pi^2 \, V'(\phi)}
  = n(\phi) \, \frac{H(\phi)}{2 \pi} ,
\end{equation}
and the width of this maximum is of the order ${H\over 2\pi}$. In
other words, most of the domains of a given field $\phi$ are formed
due to non-perturbative (hence the subscript ``np'') jumps which are
greater than the ``typical'' ones by a factor $n(\phi)$ which
coincides with our previous result (\ref{amplification}).  For future
reference, we will write here this result in an equivalent form,
\begin{equation}\label{amplification_1}
 n(\phi) = 4 \lambda_1 \frac{V(\phi)}{V'(\phi)} \ .
\end{equation}
 The limit of applicability of this expression is below the
energy level $V(\phi_{\rm fr})$ (see (\ref{phi_fr_define}),
(\ref{phi_fr_compare}) for definition of the critical value $\phi_{\rm
  fr}$).

In particular, for the theory   $\lambda \phi^4/4$  we have
\begin{equation}\label{amplification_2}
 n(\phi) =  \lambda_1 \phi  \ .
\end{equation}
For $H_{\rm max} = \sqrt{8\pi / 3}$, $\lambda \ll 1$ and $\phi \sim 4.5$,
which corresponds to today's horizon scale, this gives the amplification
coefficient
\begin{equation}\label{amplification_3}
 n(\phi) =  2\sqrt{6\pi} \phi \sim 40 \ .
\end{equation}


\section{Volume Weighted Slow Rolling Approximation}
\label{VolumeRoll}

We learned in the previous section that quantum fluctuations in volume
weighted measure have pretty large expectation value, which makes the
jumps to go preferentially downwards (unlike in comoving measure where
there is no preferred direction of the fluctuations and therefore they
have zero expectation value). Indeed, such was the very meaning of our
derivation of large jumps that they had to occur in the direction of
usual slow roll in order to make up the extra time spent by inflaton
at higher energies. Therefore, we can conclude that the slow rolling
speed itself gets a correction corresponding to the rate at which such
large jumps occur and their size. Since each such jump of the size
$n(\phi) \cdot H(\phi)/2 \pi$ occurs during time interval
$H^{-1}(\phi)$, we can estimate the additional speed gained by the
inflaton as $n(\phi) \cdot H^2(\phi)/2 \pi$, thus bringing the overall
slow roll speed to the volume weighted value (we substituted
(\ref{amplification}) for the value of $n(\phi)$, the amplification
factor):

\begin{equation}
  \label{volume_slow_roll}
  \dot \phi = -\frac{V'(\phi)}{3 \, H(\phi)}
  - d_{\rm fr} \,  H_{\rm max} \, \frac{16 V^2(\phi)}{3 V'(\phi)} \ .
\end{equation}
Here the minus sign in front of the correction term is due to the
preferred direction of the jumps, bringing the slow roll speed to
higher absolute value.

The limits of applicability of this expression are the same as for
(\ref{jump_size}), i.\ e.\ below energy density corresponding to the
critical value $\phi_{\rm fr}$ of inflaton field, defined by
(\ref{phi_fr_define}), (\ref{phi_fr_compare}). However, those limits
simply tell where the approximate expression (\ref{volume_slow_roll})
is valid, while the effect of speeding up the slow roll of the
inflaton is valid in a much wider range.

Let us derive a more general version of this result and,
correspondingly, a more general expression for amplified quantum jumps
(\ref{amplification}), (\ref{jump_size}) which will be valid for
almost whole range of variation of the inflaton field. The volume
weighted probability distribution can be defined as the path integral
over all realizations of noise taken with gaussian weight modified by
the volume factor \cite{Nambu,LLM}:

\begin{equation}
  \label{path_int_P_p}
  P_p(\phi, t) = \int {\cal D}\xi \, \exp\left\{ \int\limits^{t}
     \left(- \frac{1}{2} \xi^2(s)
      + 3 H(\phi_{\xi}(s)) \right) \, ds \right\}
  \, \delta\left(\phi_{\xi}(t) - \phi \right) \, .
\end{equation}
Here $\phi_{\xi}(s)$ is the solution of (\ref{SDE}) with a particular
realization of the noise. The gaussian path integral over the noise
can be converted into the path integral over the histories of inflaton
evolution \cite{Shtanov} if we express the noise through concurrent
value of inflaton $\phi(t)$ using the equation of motion (\ref{SDE}):

\begin{equation}
  \label{noise_phi}
  \xi(t) = \frac{2 \pi}{H^{3/2}(\phi)} \dot\phi
  + \frac{2 \pi \, V'(\phi)}{3 \, H^{5/2}(\phi)}\ .
\end{equation}
It is convenient to make the following variable transformation:

\begin{equation}
  \label{var_transform}
  z = \int\limits_{\phi} \frac{2 \pi}{H^{3/2}(\phi')} \, d \phi' \ .
\end{equation}
In terms of this variable the definition of the white noise is
rewritten in compact form:

\begin{equation}
  \label{noise_z}
  \xi(t) = - \dot z + W(z)\ ,
\end{equation}
where we introduced the ``superpotential''\footnote{This name is due
  to the fact that $W(z)$ plays a role of a superpotential in a
  SUSY-Schrodinger like version of Fokker-Planck equation.} (we used
the relation (\ref{var_transform}) to re-express it in terms of the
derivative with respect to $z$):

\begin{equation}
  \label{superpotential}
  W(z) = \frac{2 \pi}{3 \, H^{5/2}(\phi)} \,
  \frac{d \, V(\phi)}{d\, \phi}
  = \frac{d}{dz} \left(\frac{3}{16 \, V(z)} \right) \ .
\end{equation}

The path integral defining the volume weighted measure in terms of
$z(t)$ becomes, after substituting (\ref{noise_z}) into
(\ref{path_int_P_p}):

\begin{equation}
  \label{path_int_z}
    P_p(z, t) = \int {\cal D}z(s) \, J[z]
    \, \exp\left\{- \int\limits^{t}
     \left(\frac{1}{2} \left[\dot z(s) - W\left(z(s) \right) \right]^2
      - 3 H(z(s)) \right) \, ds \right\} \,
  \delta\left(z(t) - z \right) \, .
\end{equation}
The Jacobian $J[z]$ of the transformation from $\xi$ to $\phi$ and
then to $z$ is pre-exponential \cite{Shtanov} and unimportant for our
current investigation. We will neglect it in what follows.

Let us find the trajectory $z(t)$ [which we will translate later into
trajectory $\phi(t)$] which contributes most to the path integral
(\ref{path_int_z}). Such saddle point trajectory will correspond to
the typical history of evolution of inflaton under volume weighted
measure. The exponent in the path integral (\ref{path_int_z}) looks
like a Euclidean version of Lagrangian action, which corresponds to
interpretation of diffusion equation (\ref{BranchFP}) as a Euclidean
Schrodinger equation for a point particle. We can rewrite this action
in Hamiltonian form using the conventional relation:

\begin{equation}
  \label{Lagrangian_to_Hamiltonian}
  \int\limits^{t} {\cal L} \, dt = \int\limits^{z(t)} p \, dz -
  \int\limits^{t} {\cal H} \, dt \, ,
\end{equation}
where the canonical momentum is
\begin{equation}
  \label{canonical_p}
  p = \frac{\partial {\cal L}}{\partial \dot z} = \dot z - W(z) \ .
\end{equation}
Since the action does not contain explicit time dependence, the
Hamiltonian is conserved:

\begin{equation}
  \label{conserved_H}
  {\cal H} = \frac{1}{2} p^2 + p \, W(z) + 3 \, H(z) = \lambda_1\ .
\end{equation}
The reason why the conserved Hamiltonian is equal to the highest
eigenvalue is that at the end we should get the time dependence of a
type $\exp\left( \lambda_1 \, t \right)$ as warranted by the stationary
solution (\ref{Eigenseries}). Meanwhile, the $\int p \, dz$ term of
the action should give us the correct (semi-classical) field
dependence of the probability density $P_p(z(\phi), t)$ (see below).

Solving the Hamiltonian constraint (\ref{conserved_H}) with respect to
$p$ (we have to choose the positive solution of the equation for
rolling down), and using (\ref{canonical_p}), we obtain the equation
for the typical volume weighted trajectory:

\begin{equation}
  \label{volume_roll_z}
  \dot z = \sqrt{W^2(z) + 2 \lambda_1 - 6 H(z)}\ .
\end{equation}
This equation translates back in terms of inflaton field variable into
volume weighted slow roll equation:

\begin{equation}
  \label{volume_roll_phi}
  \dot \phi = - \sqrt{ \left(\frac{V'(\phi)}{3 H(\phi)} \right)^2
      +  \Bigl( d_{\rm fr} \, H_{\rm max} - 3 H(\phi)\Bigr)
   \frac{H^3(\phi)}{2 \pi^2}    }\ .
\end{equation}
For most of the inflaton range of variation (except very close to
Planck boundary) we can ignore $3 H$ term with respect to $d_{\rm fr}
H_{\rm max}$ term.  The relative importance of the two remaining terms
under the square root is governed by the critical value $\phi_{\rm
  fr}$ --- below this level the first term is more important, while
above it the second one dominates. Not surprisingly, below $\phi_{\rm
  fr}$ this equation coincides with (\ref{volume_slow_roll}). However,
its validity limits are much wider, allowing us to use it beyond
$\phi_{\rm fr}$, to which the applicability of
(\ref{volume_slow_roll}) was limited.

We can write down a good approximation for the field
dependent normalized probability density $\pi_1(\phi)$, omitting the
less important pre-exponential terms:
\begin{eqnarray}
  \label{WKB_pi_1}
 \pi_1(\phi)&=&\exp \left\{ - \int\limits^{z(\phi)} p \, dz \right\}
\nonumber\\
 &=&\exp \left\{ - \int_{\phi}{ d\zeta \, \left(
    \sqrt{ \left(\frac{3 \, V'(\zeta)}{16 \, V^2(\zeta)} \right)^2   +
    \Bigl( d_{\rm fr} \, H_{\rm max} - 3 H(\phi)\Bigr)   \frac{ 8 \pi^2
}{H^3(\phi)} } -
    \frac{3 \, V'(\zeta)}{16 \, V^2(\zeta)} \right) } \right\}. ~~~~
\end{eqnarray}
Of course, below $\phi_{\rm fr}$ this expression also coincides with
its counterpart (\ref{no_diffusion_solution}) derived previously. This
result has remarkable properties which will be studied further in
\cite{Disser}.

Using the volume weighted slow roll equation (\ref{volume_roll_phi})
we can derive a   general expression for the amplified quantum jump
size. It is given by the change in $\phi$ within time interval
$H^{-1}$ calculated according to (\ref{volume_roll_phi}), less the
regular expression for the change of field due to the slow roll in
comoving measure:

\begin{equation}
  \label{jump_size_general}
  \Delta \phi_{\rm np} = \sqrt{
    \left(\frac{V'(\phi)}{3 \, H^2(\phi)} \right)^2
      + \Bigl( d_{\rm fr} \, H_{\rm max} - 3 H(\phi)\Bigr)
   \frac{H(\phi)}{2 \pi^2}  }~ -~\frac{V'(\phi)}{3 \, H^2(\phi)}\  .
\end{equation}
This gives the  following expression for the  amplification factor (the ration
of $\Delta \phi_{\rm np} $ and the conventional amplitude $H/2 \pi$):

\begin{equation}
  \label{amplification_general}
  n(\phi) = \sqrt{
    \left(\frac{2 \pi \, V'(\phi)}{3 \, H^3(\phi)} \right)^2
      +    \frac{2d_{\rm fr}H_{\rm max}-6 H(\phi)}{H(\phi)}
      }~~ - ~\frac{2 \pi \, V'(\phi)}{3 \, H^3(\phi)}\  .
\end{equation}

The consistency conditions for our results (\ref{volume_roll_phi}) --
(\ref{amplification_general}) arise from several assumptions which we
made in their derivation and whose validity should be maintained. The
first one is that the slow rolling approximation is  valid, i.e.  $\ddot \phi
\ll 3 H(\phi) \dot \phi$. The second is that the
amplification factor is greater than one. The third condition is that
the saddle-point approximation used to derive these results is valid,
which means that $p \gg 1$. And the final, fourth condition is the
implicit assumption that large quantum jumps  which occur in a single
$h$-region do not make the gradient energy inside that region greater
than the potential energy of the inflaton field (which, of course, would
immediately invalidate the inflationary approximation). One can easily
check that all four conditions lead to the same, very relaxed
restrictions --- the energy density of the inflaton field $V(\phi)$ must be
lower than the
Planck density (or, more precisely, lower than the energy density
corresponding to the maximal rate of expansion $H_{\rm max}$). Thus,
we can use the  results obtained above in most of the variation range for the
inflaton
field in chaotic inflation.

One can easily check that for $\phi <   \phi_{\rm fr}$, $H(\phi) \ll H_{\rm
max}$  eq. (\ref{amplification_general}) yields \begin{equation}\label{strange}
n(\phi) =4d_{\rm fr} H_{\rm max} \frac{V(\phi)}{V'(\phi)} \ .
\end{equation}
This expression coincides with the expression for the amplification coefficient
which we obtained earlier by two other methods, see eqs.  (\ref{amplification})
and (\ref{jump_size}).

\section{Numerical Simulations} \label{Simul}

\subsection{The basic idea of computer simulations}

Even though we verified our results by several different methods, they are
still very unusual and   counterintuitive. Therefore we  performed a computer
simulation of stochastic processes in inflationary universe, which allows to
obtain an additional verification of our results and to calculate the
amplification factor $n(\phi)$ numerically.  We have used two different methods
of computer simulations. The first one is more direct and easy to understand.
Its basic idea  can be explained as follows.

 We have studied  a set of domains of initial size $H^{-1}$ filled with large
homogeneous field $\phi$. We considered large initial values  of $\phi$, which
leads to the self-reproduction of inflationary domains. From the point of
view of stochastic processes which we study, each domain can be modelled by a
single point with the field $\phi$ in it.  Our purpose was to study the typical
amplitude of  quantum jumps of the scalar field $\phi$ in those domains which
reached some value  $\phi_0 = O(1)$  close to the end of inflation. Then we calculate the amplification factor $n(\phi_0)$ for various $\phi_0$.

Each step of our calculations corresponds to a time change
$\Delta t = u H^{-1}_0$. Here $H_0 \equiv H(\phi_0)$,  and
$u$ is some number, $u < 1$.  The results do not depend on   $u$ if  it is
  small enough.
The evolution of the field $\phi$ in each domain consists of  several
independent parts.  First of all, the field evolves according
to classical equations of motion during inflation, which means that it
decreases by ${uV' \over 3H\, H_0}$ during each time interval $u H^{-1}_0$.
Secondly,
it makes quantum jumps by $\delta\phi ={H  \over 2\pi}
\sqrt{u H \over H_0} r_i\, $. Here $r_i $ is a
set of normal random numbers,  which are different for each inflationary
domain.
To account for the growth of physical volume of each domain we   used  the
following procedure. We followed each domain until its radius grows two times,
and after
that we considered it as 8 independent
domains.   In
accordance with  our condition $ P_p(\phi,t)|_{\phi>\phi_p} = 0$, we removed
all domains where the field $\phi$ jumped to the super-Planckian densities
$V(\phi) > 1$. Therefore our method removes the overall growth factor
$e^{\lambda_1 t}$ in the expression $P_p \sim e^{\lambda_1 t} \pi_1(\phi)$ and
directly gives  the time-independent function $\pi_1(\phi)$ which we are
looking for. Indeed we have checked that after a sufficiently large time $t$
the distribution of domains followed by the
computer with a good accuracy approached the stationary distribution
$\pi_1(\phi)$
which we have obtained in \cite{LLM} by a completely different method, see Fig.
\ref{F1}. We used it as a consistency check for our calculations. In what
follows we will not distinguish between $P_p$ and the time-independent factor
$\pi_1(\phi)$.

We kept in the computer memory information about all jumps of each domain
during the last time interval $H_0^{-1}$ before the field $\phi$ inside this
domain becomes smaller than $\phi_0$. This made it possible to evaluate an
average sum of all jumps of those domains in which the scalar field  became
smaller than   $\phi_0$ within the last time interval $H_0^{-1}$.
Naively, one could expect   this value to be smaller than ${H_0\over 2\pi}$,
since the average amplitude of the jumps is ${H_0\over 2\pi}$, but they occur
both in the positive and negative directions. However,  our simulations
confirmed our analytical result $\Delta\phi = \lambda_1 \phi   \cdot {H_0\over
2\pi}$. In other words, we have found  that most of the domains which   reach
the hypersurface  $\phi = \phi_0$ within a time interval $\Delta t = H_0^{-1}$
do it by rolling accompanied by persistent jumps down, which have a combined
amplitude $\lambda_1 \phi_0 $ times greater than ${H_0\over 2\pi}$.

\subsection{Details of the method}
Even though this method of calculations may seem quite straightforward, (it is
the so-called event-tracing Monte-Carlo method) in reality it must be somewhat
modified. The main problem is obvious if one recalls our expression for the
probability distribution $P_p \sim e^{\lambda_1 t} \pi_1(\phi)$ at small
$\phi$:\, $P_p \sim   e^{\lambda_1 t}\  \phi^{\sqrt{6\pi\over
\lambda}\lambda_1}$. As we already mentioned, (omitting the  time-dependent
factor) this yields $P_p \sim    \phi^{10^8}$ for the realistic value $\lambda
\sim 10^{-13}$. It is extremely difficult to work with distributions which are
so sharp.

Therefore in our computer simulations we have studied models with $\lambda \sim
 0.1$, which makes computations possible. On the other hand,  when one
increases the value of  $\lambda$ an additional problem arises. Our simple
expression $P_p \sim     \phi^{\sqrt{6\pi\over \lambda}\lambda_1}$ has been
obtained in the limit of very small $\lambda$, which is not perfectly accurate
for $\lambda \sim 0.1$. Therefore we will represent $P_p$ in a more general
form $P(\phi)=\phi^{g(\phi)}$, where $g(\phi)$ approaches a constant value
${\sqrt{6\pi\over \lambda}\lambda_1}$ for  $\phi\ll \lambda^{-1/8}$.  One
should also take into account that the
classical decrease $c(\phi)={V'(\phi) \over 3H\, H_0}$  of the field $\phi$
during the time $H^{-1}_0$   and the standard deviation   $s(\phi) = {H\over
2\pi}\sqrt{H\over H_0}$ (the average amplitude of quantum fluctuations during
the time $H^{-1}_0$)  are not constant throughout the region where the effect
takes place.
In such a situation an expression for the amplification coefficient $n(\phi)$
will be slightly different from our simple expression $n = \lambda_1\phi$.
Therefore we should first derive here a more accurate expression for $n(\phi)$,
and then compare it with the results of our simulations.

  Consider a point for which $\phi=\phi_0$ at some time $t$, at which the
stationary regime is already established. At the earlier time $t-H_0^{-1}$
this point was
approximately at $\phi=\phi_0+c(\phi_0)+x$, where   $x = \delta\phi$ is the sum
of all quantum jumps experienced by the field $\phi$ at this point  during the
last time interval $H_0^{-1}$. Consider the probability  $P(\delta\phi)$
that the field $\phi$ jumped to $\phi_0$ from the point $\phi_0+c(\phi_0)+x$.
This probability distribution is equal to the  distribution
$P_p(\phi)$ times the probability of undergoing a quantum fluctuation of length
$\delta\phi$.
\begin{equation}
P(x)\propto P_p(\phi_0+c(\phi_0)+x)\ \exp\left(- {x^2\over 2s^2} \right) \ .
\end{equation}
Position of the  maximum of  the distribution $P(x)$ is given by
\begin{equation}
\label{MEQ1}
{P_p'(\phi_0+c(\phi_0)+x) \over P_p(\phi_0+c(\phi_0)+x)}={x \over s^2} \ .
\end{equation}
To solve this equation for $x$ we need to know $P_p(\phi)$. As earlier we
assume
$P_p(\phi)=\phi^{g(\phi)}$ where $g(\phi)$ varies slowly with $\phi$. If
${g'\phi\ln\phi } \ll \phi^{-1}\ln g$ (which happens to be a good approximation
for $\phi \sim 1$), eq. (\ref{MEQ1}) can be easily solved, and the
expression for $n(\phi_0)$ looks as follows:
\begin{equation}
\label{MEQ2}
n(\phi_0) = {x\over s(\phi_0)} \approx {1 \over 2s(\phi_0)} \left( \sqrt{(\phi_0 +
c(\phi_0))^2+4g
(\phi_0) s^2(\phi_0)}-(\phi_0 + c(\phi_0)) \right) \ .
\end{equation}
One can obtain a slightly more accurate expression by taking into account
dependence of $g$, $c$ and $s$ on $\phi$.
Note that in the situation which we are going to investigate $\phi_0 \gg
c(\phi_0) \gg  x  \gg s(\phi_0)$. In the limit when  $g$,   $s$, and $c$ can be
considered constant, and $\phi_0 \gg s,c$ this equation leads to our earlier
expression   $n(\phi_0) = \lambda_1\phi_0$.

In order to use (\ref{MEQ2}) we also need to know $g(\phi)$ for our problem. We
approximate $g(\phi)$ by a second order polynomial in $\phi$ and substitute
$P(\phi)=\phi^{a_0+a_1\phi+a_2\phi^2}$ into the differential equation for
$P(\phi)$. Local analysis around $\phi=\phi_0$ shows $P(\phi)\approx
\phi^{56-23\phi-4\phi^2}$. This approximation is accurate for  
$\phi \sim1$.

\subsection{Numerical Calculation of $n(\phi)$}

Even for not very small $\lambda$ the distribution $P_p$ remains extremely
sharp. We have made our simulations with $\lambda = 0.1$, in which case  $P_p
\sim \phi^{60}$. This means that if we want to follow  evolution of a single
domain with $\phi = 0.5$, then we should simultaneously keep track of $2^{60}
\sim 10^{18}$ domains with $\phi = 1$.  Therefore the simple event-tracing
Monte-Carlo approach which we described above can be quite adequate for the
investigation of   $P_p$ near its maximum, but not for the study of $P_p$ far
away from the maximum of the distribution.

A more advanced approach is  to represent  the distribution by evenly
spaced points with weight proportional to the distribution. In other words, we
rewrite the probability distribution as a finite sum of nearly delta-functional
distributions,
\begin{equation}\label{division}
P_p(\phi,t)\approx \sum_{i=1}^N  p_i(\phi,t)\ ,
\end{equation}
where
\begin{equation}
p_i(\phi,t)=\cases{
P_p(\phi_i) & for $\phi_i\leq \phi<\phi_{i+1}$ \cr
0 & otherwise
}.
\end{equation}

At each step of the simulation we investigate the evolution of the
distributions  $p_i$  during the time $\Delta t = uH^{-1}_0$. The
following equation takes into account classical decrease of the field $\phi$,
quantum
fluctuations, and inflation.
\begin{equation}
p_i(\phi,t+\Delta t)\propto p_i(\phi,t)\, {1 \over s(\phi_i)}\exp\left( -
  {(\phi-\phi_i+c)^2  \over 2\,s^2(\phi_i)}  \right)\exp\left(
{3uH(\phi)\over H(\phi_0)}\right) \ .
\end{equation}
We find $P_p(\phi,t+\Delta t)$ by computing the sum of $p_i(\phi,t+\Delta t)$.
Then we normalize the distribution $P_p(\phi,t+\Delta t)$ and again subdivide
into a new set of  $p_i$, in accordance with (\ref{division}). We repeat this
process until the resulting distribution $P_p$ approaches a stationary regime.


The most tricky part of the algorithm is to find the amplification factor
$n(\phi_0)$. To do that,  we associate another distribution $x_i(\phi,t)$ with
each $\phi_i$.  Here $x_i(\phi,t)$ is the sum
of quantum fluctuations during the last time interval $H^{-1}$, along all
trajectories which ended up in the interval $\phi_i\leq \phi<\phi_{i+1}$ at the
time $t$. We combine all $x_i(\phi,t)$ into a single distribution $X(\phi,t)$,
and evolve it in the same way as
$P_p(\phi,t)$, dividing it into the   nearly delta-functional distributions
$x_i(\phi,t)$ at every iteration. This is
possible because $x_i(\phi,t)$ is approximately gaussian and its standard
deviation is small compared with its mean. When the $P_p(\phi,t)$ converges,
$x_0(\phi_0,t)$ approximates $x$, from which $n(\phi_0)$ can be calculated.

Decreasing step size $u$
increases accuracy of $P_p(\phi)$ until some point, after which the
accuracy starts to decrease. This decrease is explained by the fact
that evolved $p_i$'s are too sharp and therefore are represented
inaccurately. To avoid this, having fixed $N$, we must keep $u$ high
enough, so that the smallest quantum fluctuation $s$ is wider than the
grid spacing. Grid spacing is proportional to $1/N$ and $s$ is
proportional to $\sqrt{u}$, so the minimal $\sqrt{u}$ is proportional to
$1/N$. Execution time until convergence is proportional to
$N^2/\sqrt{u}$, or for the minimal $u$ it is proportional to $N^3$. In
practice the largest $N$ for which the algorithm converges in a
reasonable amount of time is of the order of $10^3$.

\subsection{Results of Numerical Calculations}
\begin{figure}
\centerline{ \epsfbox{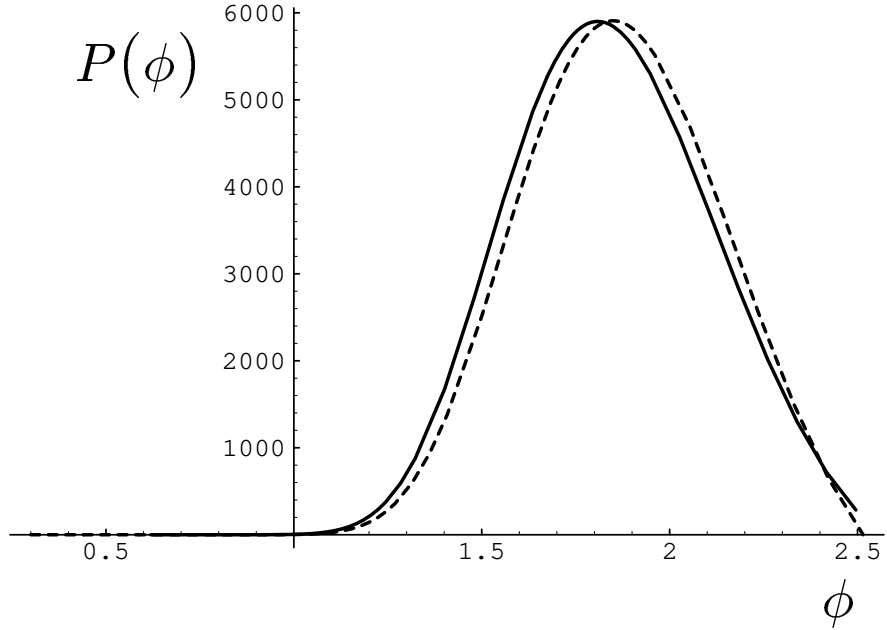}}
 \vskip 1cm
\caption{Probability distribution $P(\phi)$ for $V=\lambda \phi^4/4$,
$\lambda=0.1$. The dashed line is the numerical solution to a differential
equation describing $P(\phi)$. The solid curve is obtained using computer
simulations
described in this paper. A small deviation between the solid curve and the
dashed
line is due to the finite  size of each step and the finite grid size. }

\label{F1}

\end{figure}

The first step is to verify the numerical algorithm by comparing the
probability distribution $P(\phi)$ it
computes with a solution obtained by solving    equation (\ref{StationaryFP})
obtained in \cite{LLM}. Figure \ref{F1}
shows that $P(\phi)$ is very close to the correct probability distribution. The
deviation between the two decreases with step-size.

\begin{figure}

\centerline{ \epsfbox{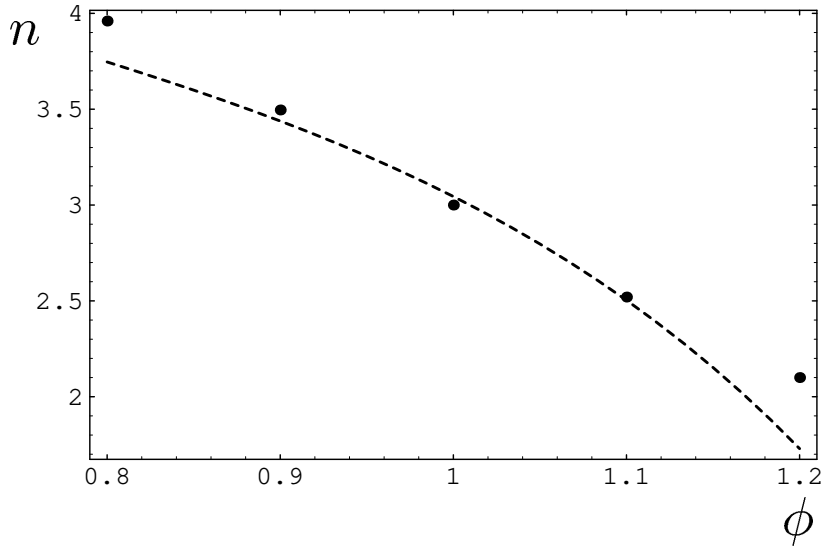}}
 \vskip 1cm
\caption{Comparison between the analytical expression for
$n(\phi)$ (dashed line) and the values for $n(\phi)$ obtained
by computer simulations.
While the analytical expression is not absolutely precise due
to various assumptions (such as constancy of $g, c, s$
during the time   $H^{-1}$), it does give
approximately correct values for $n$.}

\label{F2}

\end{figure}

The second step is to verify out formula for $n$. Figure \ref{F2} shows that
numerically computed values of $n$ for different $\phi_0$ are close to the ones
predicted by the analytical result. The deviation is explained by the
approximations made in the analytical solution (constancy of $g,c,s$ during
time $H^{-1}$). We have found also, that the typical deviation of the amplitude
of jumps from their average value $n H/2\pi$ is of the order of $H/2\pi$, as
suggested by eq. (\ref{jump_probability}). This will be important for our
subsequent considerations.

\section{The Spatial Structure of Infloids} \label{Infloids}

As one can see from eq. (\ref{volume_roll_phi}), the value of the field
$\phi(t)$ corresponding to the typical volume weighted trajectories moves down
more rapidly that one would expect from the classical slow roll equation $\dot
\phi = - \frac{V'(\phi)}{3 H(\phi)}$. This is exactly the reason why such
nonperturbatively enhanced trajectories, being surrounded by usual classical
neighbors, should correspond to the minima in the distribution of density. To
analyse   the spatial structure of the universe near the points corresponding
to the optimal volume-weighted trajectory (\ref{volume_roll_phi}) one should
remember
that in terms of the ordinary comoving measure $P_c$ the probability of large
fluctuations  is suppressed by the factor $\exp\left(-
  n^2(\phi)/2 \right)$. It is well known that exponentially suppressed
perturbations typically give rise to spherically symmetric bubbles
\cite{Adler}.
Let us show first of all that the main part of the volume of the universe in a
state with a given $\phi$ (or with a given density $\rho$) corresponds
to the centers of these bubbles,  which we called infloids.

Consider again the collection of all parts of the
universe with a given $\phi$ (or a given density)  at a given time $t$. We have
found that
most of the jumps producing this field $\phi$ during the previous time
interval $H^{-1}$ occurred from domains containing the field $\phi$ in
a narrow interval of values near $\phi -{\dot \phi}/ H + n(\phi) \cdot
H/ 2\pi$.  The width of this interval was found to be of the order of
$H/ 2\pi$, which is much smaller than the typical depth of our bubble
$\Delta\phi \sim n(\phi) \cdot H/ 2\pi$, since we have $n(\phi) \gg 1$
for all chaotic inflation models.  Now suppose that the domain
containing the field $\phi$ appears not at the center of the bubble,
but at its wall. This would mean that the field near the center of the
bubble is even smaller than $\phi$. Such a configuration could be
created by a jump from $\phi -{\dot \phi}/ H + n(\phi) \cdot H/ 2\pi$
only if the amplitude of the jump is greater than $n(\phi) \cdot H/
2\pi$.  However, we have found that the main contribution to the
volume of domains with a given $\phi$ is produced by jumps of an
amplitude $\left( n(\phi) \pm 1 \right) \cdot H/ 2\pi$, the greater
deviation from the typical amplitude $n(\phi) \cdot H/ 2\pi$ being
exponentially suppressed.  This means that the scalar field $\phi$ can
differ from its value at the center of the bubble by no more than the
usual amplitude of scalar field perturbations $H/ 2\pi$, which is
smaller than the depth of the bubble by a factor $n^{-1}(\phi)$.
Thus, the main fraction of the volume of the universe with a given
$\phi$ (or with a given density of matter) can be only slightly
outside the center. This may lead to a small contribution to anisotropy of the
microwave background radiation.

We should emphasize that all our results are based on the investigation of
the global structure of the universe rather than of the structure of
each particular bubble. This is why we assert that our effect is {\it
  non-perturbative}. If one neglects that the universe is a fractal
and looks only at one particular bubble (i.e. at the one in which we
live now), then one can find that inside each bubble there is a
plenty of space far away from its center. Therefore one could conclude
that there is nothing special about the centers of the bubbles.
However, when determining the fraction of domains near the centers we
were comparing the volumes of {\it all} regions of {\it equal} density
at equal time. Meanwhile, the density $\rho_{\rm wall}$ of matter on
the walls of a bubble is greater than the density $\rho_{\rm center}$
in its center. As we have emphasized in the discussion after eq.
(\ref{SMALLPHI}), the total volume of {\it all} domains of density
$\rho_{\rm wall}$ is greater than the total volume of all domains of
density $\rho_{\rm center}$ by the factor $({\rho_{\rm wall}/\rho_{\rm
    center}})^{3\cdot 10^7}$.  Thus, it is correct that the volume of space
outside the
center of the bubble is not smaller than the space near the center.
However, going outside the center brings us to the region of a
different density, $ \rho_{\rm wall} > \rho_{\rm center}$.  Our
results imply that one can find much more space with $\rho = \rho_{\rm
  wall}$ not at the walls of our bubble, but near the centers of {\it
  other} bubbles.

\begin{figure}
\centerline{ \epsfbox{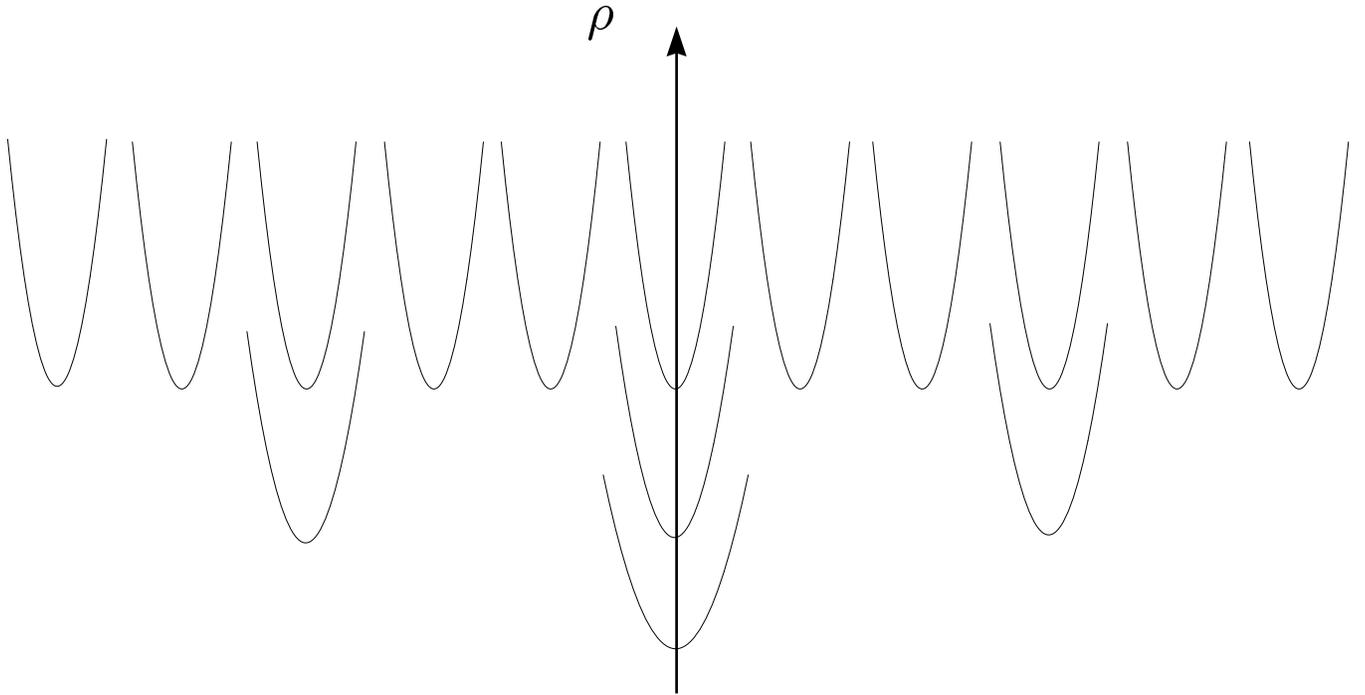}}
 \vskip 1cm
\caption{A schematic illustration which shows the number of infloids with given
density and distribution of matter near their centers.}
\label{Infloids3}
\end{figure}

This situation can be very schematically illustrated by Fig. \ref{Infloids3}.
We do not make an attempt to show the spatial distribution of
infloids. Rather we show the density distribution near the center of
each of them. All these regions basically are very similar, but at any
particular moment of time $t$ there are much more regions with large
density since they appeared from the regions which  inflated
at the nearly Planckian density for a longer time.  With time the
whole set of curves should go lower, to smaller $\rho$.  However, at
each moment of time there will be domains with all possible values of
$\rho$, so that the distribution of all curves does not change in
time (stationarity).  If one looks at the whole picture without discriminating
between states with different values of density, it may seem that there is
much more space outside of the centers of the bubbles. However, at any
given moment of time $t$ the main fraction of volume of the universe
in a state {\it with a given density $\rho$} is concentrated near the centers
of spherically symmetric bubbles. One may look, for example, at the
density corresponding to the centers of the third row of curves. At
this density one may live either near the center of any of the eleven
infloids, or at the walls of only three of them.  The fraction of the
volume near the centers would be much greater if we try to show the
realistic distribution $P_p(\rho) \sim \rho^{3\cdot10^7}$ of the number of
domains with a given density in the theory $\lambda\phi^4/4$.

The nonperturbative jumps down should occur on all scales
independently. One may visualize the whole process as follows. At each given
moment most of the volume of the universe where the field $\phi$ takes some
particular value appear close to the centers of  infloids created by the
nonperturbative jumps by $n(\phi) H/2\pi$. The new jumps occur each time
$H^{-1}$ independently of the previous history of the regions with a given
$\phi$. Therefore the leading contribution to the volume with be given by those
rare centers of infloids where the field $\phi$ jumps down by  $n(\phi) H/2\pi$
again and again.  That is why the typical volume weighted trajectories
permanently go down with the speed exceeding the speed of classical rolling by
$n(\phi) H^2/2\pi$, see eqs. (\ref{volume_roll_phi}),
(\ref{amplification_general}).

One may visualize the resulting
distribution of the scalar field in the following way.  At some scale
$r$ the deviation of the field $\phi$ from homogeneity can be
approximately represented as a well of a radius $r$ with the depth
$n(\phi) {H\over 2\pi}$. Near the bottom of this well there
is another well of a smaller radius $e^{-1}r$ and approximately of the
same depth $n(\phi) {H\over 2\pi}$. Near the center of this
well there is another well of a radius $ e^{-2}r$, etc. In particular, in the
theory $\lambda\phi^4/4$ the depth of each well will be ${3H_{\rm max} H\phi
\over 2\pi}$.  Of course,
this is just a discrete model. The shape of the smooth distribution of
the scalar field is determined by the equation

\begin{equation}\label{ScalarShape1}
 {d\phi\over d\ln rH} =   {3H_{\rm max}
H\phi\over 2\pi} = \sqrt{ 3\lambda \over 2 \pi}\,  H_{\rm max} \phi^3
\, ,
\end{equation}
 which gives

\begin{equation}\label{ScalarShape2}
 \phi ^2(r) \approx {\phi ^2(0)
\over 1 -  H_{\rm max} \, \phi^{2}(0) \sqrt{ 6\lambda\over \pi }\, \ln
{rH}} ~~~~~~ \, \mbox{for} \ \ r > H^{-1} \ {}.
\end{equation}
 Note that $\phi(r) \approx \phi(0)$ for $r < H^{-1}$
(there are no perturbations of the classical field on this scale).

This distribution is slightly altered by the usual small perturbations
of the scalar field. At a distance much greater than their wavelength
from the center of the well these perturbations have the usual
magnitude ${H\over 2\pi}$.  Thus, our results do not lead to
considerable modifications of the usual density perturbations which
lead to galaxy formation. However, the presence of the deep well
(\ref{ScalarShape2}) can significantly change the local geometry of
the universe.

 In the inflationary scenario with $V(\phi) =
{\lambda\over 4} \phi^4$ fluctuations which presently have the scale
comparable with the horizon radius $r_h \sim 10^{28}$ cm have been
formed at $\phi~\sim~5$ (in the units $M_p = 1$).  As we have
mentioned already $3H_{\rm max} \approx 2\sqrt{6\pi} \sim 8.68$ for our
choice of boundary conditions \cite{LLM}, and  the typical nonperturbative jump
down on the scale of the present horizon should be $3H_{\rm max} \phi
\sim 40$ times greater than the standard jump, see eq. (\ref{amplification_3}).
In the theory
${\lambda\over 4} \phi^4$ the standard jumps lead to density
perturbations of the amplitude ${\delta\rho\over \rho} \sim
{2\sqrt{6\lambda\pi}\over 5}\, \phi^3 \sim 5\cdot10^{-5}$ (in the
normalization of \cite{MyBook}). Thus, according to our analysis, the
nonperturbative decrease of density on each length scale different
from the previous one by the factor $e$ should be about
${\delta\rho\over \rho} \sim  H_{\rm max} {6\sqrt{6\lambda\pi}\over 5}
\phi^4 \sim 2\cdot10^{-3}$.  This allows one to evaluate the shape of
the resulting well in the density distribution as a function of the
distance from its center. One can write the following equation for the
scale dependence of density:

\begin{equation}\label{shape1}
 {1\over \rho} {d\rho\over d\ln {r\over
r_0}} = -  H_{\rm max} \cdot {6\sqrt{6\lambda\pi}\over 5}\, \phi^4 \ ,
\end{equation}
 where $r$ is the distance from the center of the
well. Note that $\phi = {1\over\sqrt \pi}\,(\ln {r\over r_0})^{1\over
2}$ in the theory ${\lambda \over 4}\phi^4$ \cite{MyBook}. Here $ r_0$
corresponds to the smallest scale at which inflationary perturbations
have been produced. This scale is model-dependent, but typically at
present it is about 1 cm. This yields

\begin{equation}\label{shape2}
 {\Delta\rho\over \rho_c} \equiv
{\rho(r) - \rho(r_0)\over \rho(r_0)} = {2
H_{\rm max}\sqrt{6\lambda}\over 5\pi\sqrt{3\pi}} \, \ln^3{r\over r_0} \
{}.
\end{equation}
 This gives the typical deviation of the density on the
scale of the horizon (where $\ln{r_h\over r_0}\sim 60$) from the
density at the center: ${\Delta\rho\over \rho_c} \sim 750 \cdot
{\delta\rho\over \rho} \sim 4\cdot 10^{-2}$.  \vskip 0.3cm

  It is very tempting to interpret this effect in such
a way that the universe around us becomes locally open, with $1 -
\Omega \sim 10^{-1}$.  Indeed, our effect is very similar to the one
discussed in \cite{ColemanDL,LMomega}, where it was shown that the
universe becomes open if it is contained in the interior of a bubble
created by the $O(4)$ symmetric tunneling.  Our nonperturbative jumps
look very similar to tunneling with the bubble formation.  However,
unlike in the case considered in \cite{ColemanDL,LMomega}, our bubbles
appear on all length scales.

The results discussed above refer to the density distribution at the
moment when the corresponding wavelengths were entering horizon. At
the later stages gravitational instability should lead to growth of
the corresponding density perturbations. Indeed, we know that density
perturbations on the galaxy scale have grown more than $10^4$ times in
the linear growth regime until they reached the amplitude
${\delta\rho\over \rho} \sim 1$, and then continued growing even
further. The same can be expected in our case, but even in a more
dramatic way since our ``density perturbations" on all scales are
much greater than the usual density perturbations which are
responsible for galaxy formation. This would make the center of the
well very deep; its density should be many orders of magnitude smaller
than the density of the universe on the scale of horizon. This is not what we
see around.

This problem can be easily resolved. Indeed, our effect (but not the
amplitude of the usual density perturbations) is proportional to
$H_{\rm max}$, which is  the maximal value of the Hubble
constant compatible with inflation. If, for example, the maximal
energy scale in quantum gravity or in string theory is given not by
$10^{19}$ GeV, but by $10^{18}$ GeV, then the parameter $H_{\rm max}$
will decrease by a factor $10^{-2}$.  As we already mentioned, even
smaller nonperturbative effects are expected in new inflation where $H_{\rm
max}$ is
always many orders of magnitude smaller than $1$. Inflationary Brans-Dicke
cosmology in cases when the probability distribution $P_p$ is
stationary also leads to   negligibly small nonperturbative effects \cite{BL}.
 Thus it is easy to
make our effect very small without disturbing the standard predictions
of inflationary cosmology. However, it is quite possible that we will
not have any difficulties even with  very large $n(\phi)$ if we interpret
our results more carefully.

\section{Interpretation and possible improvements of the probability
measure} \label{Interpretation}

An implicit hypothesis behind our interpretation is that we are
typical, and therefore we live and make observations in those parts of
the universe where most other people do.  One may argue that the total
number of observers which can live in domains with given properties
(e.g. in domains with a given density) should be proportional to the
total volume of these domains at a given time.  However, our existence
is determined not only by the local density of the universe but by the
possibility for life to evolve for about 5 billion years on a planet
of our type in a vicinity of a star of the type of the Sun. If, for
example, we have density $10^{-29} \, {\rm g} \cdot {\rm cm}^{-3}$ in a small
vicinity of   the
center of the infloid, and density $10^{-27} \, {\rm g} \cdot {\rm cm}^{-3}$ on
the
horizon scale, then the age of our part of the universe (or, to be more
accurate, the time after the end of inflation) will be determined not by the
density near the center of the infloid,  by
the   large-scale density $10^{-27} \, {\rm g} \cdot {\rm cm}^{-3}$, and it
will be
only about one billion years.

Moreover,  any structures such as galaxies or clusters cannot be formed near
the
centers of the infloids since the density there is very small.  Indeed, on each
particular scale the jump down
completely overwhelms the amplitude of usual density
perturbations. The bubble cannot contain any galaxies at the distance
from the center comparable with the galaxy scale, it cannot contain
any clusters at the distance comparable with the size of a cluster,
etc. In other words,
the center would be devoid of any structures necessary for the
existence of our life.

Thus, the naive idea that the number of observers is proportional to
volume does not work at the distances from the centers which are
smaller than the present size of the horizon. Even though at any given
moment of time most of the volume of the universe at the density
$10^{-29} \, {\rm g} \cdot {\rm cm}^{-3}$ is concentrated near the centers of infloids, the
corresponding parts of the universe are too young and do not have any
structures necessary for our existence.  Volume alone does not mean
much. We live on the surface of the Earth even though the volume of
empty space around us is incomparably greater.

One may argue,   that the disparity between the age of the local part of the
universe and its density appears only if one considers perturbations on a scale
smaller than the horizon. Therefore it still may be true that we should live in
the centers of huge bubbles, which have a shape (\ref{shape2}) for $r >
H_0^{-1}$, where $H_0^{-1}$ is the size of the present horizon. If the cut-off
occurs at $r \gg H_0^{-1}$,  this may not lead to any observable consequences
at all. However, if the cut-off occurs at $r \sim H_0^{-1}$, the resulting
geometry may resemble an open universe with a
scale-dependent effective parameter $\Omega(r)$ \cite{LLMcenter}.

In order to make any definite conclusions about the preferable parts of the
universe one should study probability distributions which include several other
factors in addition to density. This should be a subject of a separate
investigation.  An additional ambiguity in   interpretation of our results appears due to
the
dependence of the distribution $P_p$ on the choice of time
parametrization. Indeed, there are many different ways to define
``time'' in general relativity. If, for example, one measures time not
by clock but by rulers and determines time by the degree of a local
expansion of the universe, then in this ``time'' the rate of expansion
of the universe does not depend on its density. As a result, our
effect is absent in this time parametrization \cite{LLMcenter}. The reason why
the results depend on the time parametrization is deeply related to the
properties of a self-reproducing universe. The total volume of all parts of
such a universe diverges in the large time limit. Therefore when we are trying
to find which parts of the universe have greater volume we are comparing
infinities. There are some methods to regularize these infinities in a way that
would make the final results only mildly dependent on the choice of time
parametrization \cite{Vil_predict_1,LMregul}. However, there are many such
methods, and the final results are exponentially sensitive to the choice of the
method \cite{LMregul}.  In this
paper we used the standard time parametrization which is most closely
related to our own nature (time measured by number of oscillations
rather than by the distance to the nearby galaxies). But maybe we
should use another time parametrization, see Appendix, or even integrate over all
possible time parametrizations?  Right now we still do not know what is
the right way to go. We do not even know if it is right that we are
typical and that we should live in domains of the greatest volume, see
the discussion of this problem in \cite{LLM,LMregul,GBLinde}.

Therefore at present we would prefer to consider our results simply as a
demonstration of nontrivial properties of the hypersurface of a given time in
the fractal
self-reproducing universe, without making any far-reaching
conclusions concerning the structure of our own part of the universe.
However,  we must admit that we are amazed by the  fact that  the main fraction
of volume of inflationary universe
in a state with a given density $\rho$ at any given moment of proper
time $t$ should be concentrated near the centers of deep spherically symmetric
wells. We confirmed this result by four different methods, and we believe that
it is correct. Until the interpretation problem is resolved, it will remain
unclear
whether our result is just a mathematical curiosity, or it can be considered as
a real prediction of
properties of our own part of the universe. At present we can neither prove nor
disprove the last possibility, and this by itself is a very
unexpected conclusion. Few years ago we would say that the possibility
that we live in a local ``center of the world'' definitely contradicts basic
principles of cosmology. Now we can only say that it is an open
question to be studied both theoretically and experimentally.  If somebody
asks whether  we should live in the  center of the world, we will be unable to
give a definite answer.  But if observations   show us that the answer is yes,
we will know why.

\subsection*{Acknowledgements}

The authors are   grateful to   J. Garc\'{\i}a--Bellido,  V. Mukhanov,   and A.
Vilenkin for many valuable
discussions.  This work was
supported in part  by NSF grant PHY-8612280.

\section*{Appendix}\appendix
Let us consider a different time parametrization, related to the proper time
by local path dependent transformation:

\begin{equation} \label{time_reparametrization}
          t \rightarrow \tau (t) = \int^{t} ds \, T(\phi_{\xi} (s))\ ,
\end{equation}
where $T(\phi)$ is a positive function, and its argument in
(\ref{time_reparametrization}) is a solution of (\ref{SDE})
with a particular realization of the white noise. The stochastic Langevin
equation in this parametrization looks like follows:

\begin{equation} \label{SDE_any_time}
   \frac{d \phi}{d\tau}  =  -\frac{V'(\phi )}{3H(\phi ) \, T(\phi)} +
           \frac{H^{3/2}(\phi )}{2 \pi \, T^{1/2}(\phi)} \, \xi
           (\tau)
\end{equation}

The branching diffusion equation in arbitrary time parametrization can be
written as:

\begin{eqnarray}\label{BranchFP_any_time}
\frac{\partial}{\partial \tau} P_p(\phi, \tau) & = &
         \frac{1}{2} \, \frac{\partial}{\partial \phi} \left(
\frac{H^{3/2}(\phi )}{2 \pi \, T^{1/2}(\phi)}
                \frac{\partial}{\partial \phi} \left(\frac{H^{3/2}(\phi )}{2
\pi \, T^{1/2}(\phi)}
                P_p(\phi,
\tau) \right) \right) \nonumber \\
       & + & \frac{\partial}{\partial \phi} \left(\frac{V'(\phi )}{3H(\phi ) \,
T(\phi)} P_p(\phi, \tau)
\right) +  \frac{3H(\phi)}{T(\phi)} P_p(\phi, \tau)\ .
\end{eqnarray}

Its solution will generally be a stationary probability function with an
overall constant expansion factor just like in (\ref{Eigenseries}). The value
of the constant $\lambda_1$ will depend on the parametrization.

We can find the volume weighted slow roll trajectory of the inflaton field in
arbitrary parametrization very similarly to the approach used for proper time,
but have to keep in mind that it is no longer true that $\lambda_1 = d_{\rm fr}
H_{\rm max}$. The result is:

\begin{equation}
  \label{volume_roll_any_time}
  \frac{d \phi}{d \tau} = - \sqrt{ \left(\frac{V'(\phi)}{3 H(\phi) \, T(\phi)}
\right)^2
      +  \left( \lambda_1 - 3 \frac{H(\phi)}{T(\phi)}\right)
   \frac{H^3(\phi)}{2 \pi^2 \, T(\phi)}    }\ .
\end{equation}

Since the conventional (i.e.\ calculated under the comoving probability)
amplitude of the quantum jumps generated during the typical time interval
$\Delta \tau \sim T(\phi) H^{-1}(\phi)$ in the given time parametrization is
still given by the usual quantity $H/2 \pi$ (see the Langevin equation
(\ref{SDE_any_time}) above) then the definition for amplification factor
becomes:

\begin{equation}\label{amplification_any_time}
  n(\phi) =  \sqrt{ \left(\frac{2 \pi V'(\phi)}{3 H^3(\phi)} \right)^2
      +  \left( \lambda_1 - 3 \frac{H(\phi)}{T(\phi)}\right)
   \frac{2 T(\phi)}{H(\phi)}    }~ -~ \frac{2 \pi V'(\phi)}{3 H^3(\phi)} \ .
\end{equation}

In the particular case of the time parametrization $T=H$, which corresponds to
the scale factor $a(t)$ playing the role of time $\tau$, we get:

\begin{equation}
  \label{volume_roll_tau}
  \frac{d \phi}{d \tau} = -~ \sqrt{ \left(\frac{V'(\phi)}{3 H^2(\phi)} \right)^2
      +  \left( \lambda_1 - 3 \right)
   \frac{H^2(\phi)}{2 \pi^2 }    } \ .
\end{equation}
and
\begin{equation}\label{amplification_tau}
  n(\phi) =  \sqrt{ \left(\frac{2 \pi V'(\phi)}{3 H^3(\phi)} \right)^2
      +  2 \left( \lambda_1 - 3 \right)}~ -~ \frac{2 \pi V'(\phi)}{3 H^3(\phi)} \
{}.
\end{equation}

Since $\lambda_1 < 3$, in this time parametrization the volume weighted slow
roll (\ref{volume_roll_tau}) is not faster but slightly slower than the
conventional slow roll. As a result, most of the volume on the hypersurfaces of
constant ``time'' $\tau$ will be concentrated near the spherically symmetric
hills (rather than wells) in energy density. However, the amplification factor
is always very small.

The change of time parametrization  (\ref{time_reparametrization}) corresponds to one of the possible ways to choose regularization procedure for evaluation of divergent probabilities in eternally expanding universe  \cite{LMregul}. Other types of regularization procedure were proposed in    \cite{Vil_predict_1,LMregul}.   In particular,  the 
regularization scheme suggested in  \cite{Vil_predict_1} is essentially equivalent to choosing the $T=H$ parametrization  which we discussed above \cite{LMregul}.  One can easily verify that in the limit $\phi \ll \phi_{\rm fr}$
  our equations for the $T = H$ parametrization (\ref{volume_roll_tau}), (\ref{amplification_tau}) yield the same results for the nonperturbative jumps as the ones obtained in  \cite{Vil_predict_1}. As it is argued in \cite{LMregul}, from the point of view of interpretation of our results  it is not obvious that   this regularization has any advantages as
compared to a more intuitive and straightforward approach used in the main part
of this paper. However, each regularization scheme and each time parametrization   gives   an additional interesting information about the structure of inflationary universe. Therefore we presented in this Appendix an extension  of our results for the   more general class of time parametrizations (\ref{amplification_any_time}).

\newpage

\end{document}